\documentclass[aps,prd,twocolumn,showpacs,amsmath,amssymb,preprintnumbers,10pt,footinbib,superscriptaddress]{revtex4-1}
\usepackage{CJKutf8}
\usepackage{blindtext}
\usepackage{epsfig}
\usepackage{amsfonts}
\usepackage{amssymb}
\usepackage{esint} 
\usepackage{enumitem}
\usepackage{amsthm}
\usepackage{subfig}
\usepackage{bm}
\usepackage{hyperref}
\usepackage{mathrsfs}
\usepackage{bbm}
\usepackage{cancel}
\usepackage[dvipsnames]{xcolor}
\usepackage{accents}
\usepackage{relsize}
\usepackage{float, slashed, graphicx, amssymb, amsmath}
\usepackage{shuffle}
\usepackage{tikz}
\usepackage{mathtools} 
\usepackage{pgfplots}

\usetikzlibrary{hobby,calc,shapes, positioning, backgrounds,trees, decorations, decorations.markings, decorations.pathmorphing, arrows, shapes.geometric, snakes}
\usepackage{tikz-feynman}
\tikzfeynmanset{compat=1.1.0}
\tikzfeynmanset{graviton/.style={circle, draw=green!60, fill=green!5, very thick, minimum size=7mm}}
\tikzfeynmanset{codot/.style={/tikz/shape=circle,/tikz/fill=white,/tikz/minimum size=0.1cm,/tikz/inner sep=1.8pt}}
\tikzfeynmanset{myblob/.style={/tikz/shape=rectangle,/tikz/fill=red,/tikz/minimum size=0.2cm,/tikz/inner sep=1.8pt} }
\tikzfeynmanset{ghc/.style={/tikz/shape=circle,/tikz/fill=white,/tikz/minimum size=0.05cm,} }
\tikzfeynmanset{HV/.style={/tikz/shape=circle,/tikz/fill={rgb:black,1;white,2},/tikz/minimum size=0.1cm,} }
\tikzfeynmanset{GR/.style={/tikz/shape=ellipse,/tikz/fill={rgb:black,1;white,2},/tikz/minimum size=0.3cm,} }
\tikzfeynmanset{GR2/.style={/tikz/shape=ellipse,/tikz/fill={rgb:black,1;white,2},/tikz/minimum width = 1.2cm, 
    /tikz/minimum height = 3.4cm} }
\tikzset{box/.pic={\filldraw[fill=black]  (0,0) circle (2.5pt); \filldraw [fill=black] (0.5,0) circle (2.5pt); \draw [line width=5pt] (0,0) -- (0.5,0);}}
\tikzfeynmanset{LP/.style={/tikz/shape=circle,/tikz/draw={rgb:black,2;white,1}, /tikz/minimum size=0.3cm} }

\usepackage[T1]{fontenc} 

\makeatletter
\newcommand \UPlus {\mathop {\operator@font \uplus }\limits }
\makeatother
\makeatletter
\newcommand \Bigcup {\mathop {\operator@font \bigcup }\limits }
\makeatother
  \def\LabelNote#1{}
 \def\Label#1{\label{#1}%
  \smash{\hbox to0pt{\raise1ex\hbox{\tiny[#1]}\hss}}}
  
  \def\Cdot{{\cdot}}
  
\def\nn{\nonumber}

\newcommand{\blue}{\color{blue}}

\def\spa#1.#2{\left\langle#1\,#2\right\rangle}
\def\spb#1.#2{\left[#1\,#2\right]}

\def\be{\begin{equation}}
\def\ee{\end{equation}}
\def\bea{\begin{eqnarray}}
\def\eea{\end{eqnarray}}  
\allowdisplaybreaks

\newcommand{\vareps}{\varepsilon}

\newcommand{\pb}{\bar p}

\newcommand{\tF}{\widetilde F}

\def\mdot{{\cdot}}

\def\CI{{\cal I}}

\def\CY{{\cal Y}}

\begin{document} 
\begin{CJK*}{UTF8}{mj}

\preprint{ SNUTP24-003
}

\title{
Systematic integral evaluation for spin-resummed binary dynamics 
}
\author{Gang Chen}
\email{gang.chen@nbi.ku.dk}
\affiliation{Niels Bohr International Academy,
Niels Bohr Institute, University of Copenhagen,\\
Blegdamsvej 17, DK-2100 Copenhagen \O, Denmark}
\author{Jung-Wook Kim (김정욱)}
\email{jung-wook.kim@aei.mpg.de}
\affiliation{Max Planck Institute for Gravitational Physics (Albert Einstein Institute),\\
Am M\"uhlenberg 1, D-14476 Potsdam, Germany}
\author{Tianheng Wang}
\email{tianhengwang@snu.ac.kr}
\affiliation{Center for Theoretical Physics, Seoul National University, \\
1 Gwana-ro, Gwanak-gu, 08826, Seoul, South Korea}

\begin{abstract}
Computation of spin-resummed observables in post-Minkowskian dynamics typically involve evaluation of Feynman integrals deformed by an exponential factor, where the exponent is a linear sum of the momenta being integrated. Such integrals can be viewed as tensor integral generating functions, which provide alternative approaches to tensor reduction of Feynman integrals. We develop a systematic method to evaluate tensor integral generating functions using \emph{conventional} multiloop integration techniques. The spin-resummed aligned-spin eikonal at second post-Minkowskian order is considered as a phenomenologically relevant example where evaluation of tensor integral generating functions is necessary.
\end{abstract} 

\keywords{Scattering amplitudifferential equations, heavy-mass effective theory, gravitational waves, classical spin}

\maketitle
\end{CJK*}

\section{Introduction}
Inspired by quantum field theoretic approaches to gravity~\cite{Iwasaki:1971vb,Donoghue:1994dn,Bjerrum-Bohr:2002gqz,Holstein:2004dn,Holstein:2008sx,Neill:2013wsa,Bjerrum-Bohr:2013bxa,Damour:2017zjx}, modern methods to compute scattering amplitude have been widely applied to the study of classical dynamics of binary black hole scattering, from conservative dynamics~\cite{Bjerrum-Bohr:2018xdl,Cheung:2018wkq,Cristofoli:2019neg,Bern:2019nnu,Bern:2019crd,Parra-Martinez:2020dzs,DiVecchia:2020ymx,DiVecchia:2021ndb,DiVecchia:2021bdo,Herrmann:2021tct,Bjerrum-Bohr:2021vuf,Bjerrum-Bohr:2021din,Damgaard:2021ipf,Brandhuber:2021eyq,Bjerrum-Bohr:2021wwt,Bjerrum-Bohr:2022ows,Bern:2021yeh,Bern:2022jvn,Bern:2021dqo,Damgaard:2023ttc,DiVecchia:2023frv,Heissenberg:2023uvo,Adamo:2022ooq} and gravitational bremsstrahlung~\cite{Kosower:2018adc,Brandhuber:2023hhy,Herderschee:2023fxh,Elkhidir:2023dco,Georgoudis:2023lgf,Caron-Huot:2023vxl,Bohnenblust:2023qmy,Bini:2023fiz,Georgoudis:2023eke,Adamo:2024oxy,Bini:2024rsy,Alessio:2024wmz,Georgoudis:2024pdz,Brunello:2024ibk} to effects of spin~\cite{Guevara:2017csg,Arkani-Hamed:2017jhn,Guevara:2018wpp,Chung:2018kqs,Guevara:2019fsj,Arkani-Hamed:2019ymq,Aoude:2020onz,Chung:2020rrz,Guevara:2020xjx,Chen:2021kxt,Kosmopoulos:2021zoq,Chiodaroli:2021eug,Bautista:2021wfy,Cangemi:2022bew,Ochirov:2022nqz,Damgaard:2019lfh,Bern:2020buy,Comberiati:2022ldk,Maybee:2019jus,Haddad:2021znf,Chen:2022clh,Menezes:2022tcs,FebresCordero:2022jts,Alessio:2022kwv,Bern:2022kto,Aoude:2022thd,Aoude:2022trd,Alessio:2023kgf,Aoude:2023vdk,Bautista:2023szu,DeAngelis:2023lvf,Brandhuber:2023hhl,Aoude:2023dui,Riva:2022fru,Luna:2023uwd} (see also refs.~\cite{Antonelli:2019ytb,Damour:2020tta,Kalin:2020fhe,Dlapa:2021npj,Dlapa:2021vgp,Dlapa:2022lmu,Jakobsen:2023ndj,Jakobsen:2021smu,Mougiakakos:2021ckm,Jakobsen:2021lvp,Mogull:2020sak,Jakobsen:2021zvh,Vines:2017hyw,Vines:2018gqi,Liu:2021zxr,Jakobsen:2022fcj,Damgaard:2022jem,Bianchi:2023lrg,Gonzo:2024zxo} for related approaches), motivated in part by the detection of gravitational waves by the LIGO-Virgo collaboration~\cite{LIGOScientific:2016aoc}. The relevant perturbative expansion is the post-Minkowskian (PM) expansion, where general relativistic corrections are added to the unperturbed special relativistic kinematics. The study of PM dynamics is more than a theoretical interest, as they can be used in building waveform models for gravitational wave detection~\cite{Buonanno:2024byg}.

The typical integrals encountered in the study of PM dynamics can be formulated as Feynman integrals, which can be handled efficiently by modern amplitude techniques such as the integration-by-parts (IBP) reduction~\cite{Smirnov:2008iw,Lee:2014ioa,Larsen:2015ped} and the method of differential equations~\cite{gehrmann2000differential,Kotikov:1990kg,Henn:2013pwa}. The function space of the master integrals and their associated graph topologies were studied for the conservative scattering dynamics at 4PM and 5PM orders~\cite{Bern:2021yeh,Dlapa:2021npj,Damgaard:2023ttc,Frellesvig:2023bbf,Driesse:2024xad,Klemm:2024wtd,Frellesvig:2024zph}. The function space remains practically the same when spin effects are incorporated perturbatively into the non-spinning dynamics. 

However, the function space becomes completely different when \emph{exact} spin dependence\textemdash or spin resummation\textemdash is considered. The motivation for studying spin resummation in binary Kerr dynamics is that gravitational wave physics is becoming a precision science; one potential obstruction for high-precision gravitational wave physics is inaccurate modelling of spin effects in gravitational waveform models~\cite{Dhani:2024jja}, which needs to be improved with better understanding of spin effects in gravitational two-body dynamics. For example, the waveform models based on the effective-one-body formalism~\cite{Buonanno:1998gg} currently employed by the LIGO-Virgo-KAGRA collaboration (\texttt{SEOBNRv5} family)~\cite{Pompili:2023tna,Ramos-Buades:2023ehm,Khalil:2023kep} are known to perform worse when black holes' spins approach extremality. We can expect to improve their performance in those regions of the parameter space by studying analytic properties of spin resummation and incorporating them into how perturbative two-body dynamics are resummed~\cite{Kim:2024grz}.

The spin-induced multipole moments of black holes can be generated by the Newman-Janis shift~\cite{Newman:1965tw,Arkani-Hamed:2019ymq}, where the position of the black hole is complexified and shifted in the imaginary spin direction. The translation operator in the imaginary direction is $\exp( i \alpha \cdot \mathbf{\nabla} ) = \exp( \alpha \cdot K )$, and the same factor appears when the Newman-Janis shift is promoted to a dynamical statement~\cite{Kim:2024grz}. In particular, the typical Feynman integrals encountered in the context of PM dynamics are modified by exponential factors corresponding to translations along the imaginary spin direction.

Such integrals can be considered in a more general setting as \emph{tensor integral generating functions} (TIGFs), where a typical Feynman integrand is deformed by an exponential factor of integration variables. The TIGFs provide an alternative route to tensor reduction of Feynman integrals~\cite{Feng:2022hyg}. More importantly, consideration of such integrals allows us to treat the impact parameter space Fourier transform on an equal footing with loop momenta integration, and leads us to new insights on PM dynamics presented in impact parameter space~\cite{Brunello:2024ibk}. For example, the spurious poles from the intermediate loop integral reduction, and computation of the terms that contribute to the loop integrals but vanish under the impact parameter space transform, can be avoided by performing IBP reduction at the full integrand level. 

Unfortunately, known multiloop integration techniques are not directly applicable to evaluation of such modified Feynman integrals. We introduce a method to tame the exponential factor and convert it into an extra delta constraint, rendering the integrand suitable for application of \emph{conventional} multiloop integration techniques.  Interestingly, the extra delta constraint leads to significantly different function space of the master integrals when compared to the original master integrals with the same topology. 

As a phenomenologically important application, the developed method is used to compute the 2PM spin-resummed eikonal for binary Kerr scattering. To simplify the analysis we consider a generalisation of the aligned spin configuration, where the spin vectors are aligned but allowed to have components along the impact parameter. The minimal coupling three-point amplitude~\cite{Arkani-Hamed:2017jhn} and the spin-resummed heavy-mass effective field theory (HEFT) Compton amplitude~\cite{Bjerrum-Bohr:2023iey,Bjerrum-Bohr:2023jau} are used to build the integrand from unitarity cuts \cite{Bern:1994zx,Bern:1994cg,Bjerrum-Bohr:2002gqz} and heavy-mass/velocity cuts \cite{Brandhuber:2021eyq,Bjerrum-Bohr:2021vuf,Bjerrum-Bohr:2021din}. 

\section{General methods for tensor integral generating functions} \label{sec:gen_met}
TIGFs are obtained from a typical loop integrand as a deformation by an exponential factor, viz.~\footnote{Although generating functions are usually defined by real exponents, imaginary exponents were used to avoid factors of $i$ when solving for the integrals. Such an analytic continuation is known in the probability literature as characteristic functions. This analytic continuation is allowed since Feynman integrals should be understood as generalised functions which may not be well-defined by conventional definitions of integrals.}
\begin{align}
    \mathcal{I}^{(\alpha)}[\mathbf{y}]:= \int \prod_{j=1}^L d^D K_j \frac{e^{i(\sum_{j=1}^L\alpha_j \cdot K_j)}\Big(\prod\limits_{k=r+1}^n\delta^{\lambda_{k}{-}1}(\mathcal{D}_{k})\Big)}{ \mathcal{D}_1^{\lambda_1} \cdots \mathcal{D}_r^{\lambda_r}} \,, \label{eq:TIGF_def}
\end{align}
where $\delta^{\lambda}(x) = \frac{d^{\lambda}}{dx^{\lambda}} \delta(x)$, $(\alpha) = (\alpha_1^\mu , \cdots , \alpha_L^\mu)$ is the vector of arguments, $\mathbf{y}$ is the vector of kinematic invariants and $\mathcal{D}_i$ are the inverse propagators; $(K + p)^2 - m^2$ or $2 (K \cdot p)$ when eikonalised~\footnote{The delta constraints can be converted to denominators through the distributional identity $2 \pi i \, \delta(x) = (x - i0^+)^{-1} - (x + i0^+)^{-1}$ and its derivatives, therefore they can be considered as propagator factors.}.  In the rest of the letter, we set $D=4-2\epsilon$ for dimensional regularisation. Integrals with non-trivial numerator factors can be obtained from TIGFs as derivative operators of $\alpha_j^{\mu}$ acting on them, which is a less-explored tensor reduction method for Feynman integrals~\cite{Feng:2022hyg}. This alternative tensor reduction approach may provide more efficient methods to reduce irreducible numerators, which are one of the bottlenecks in evaluating Feynman integrals.

Some specialised techniques were developed for their reduction and evaluation, including modification of  IBP reduction to incorporate the extra exponential factor~\cite{Feng:2022hyg,Hu:2023mgc,Feng:2024qsa,Li:2024rvo,Brunello:2023fef,Brunello:2024ibk}. We take a different route and develop a systematic method to reduce and evaluate TIGFs using \emph{conventional} multiloop integration techniques. Our key proposal is converting TIGF integrands \eqref{eq:TIGF_def} into a typical Feynman integrand; we introduce an auxiliary parameter $t$ and Fourier transform the exponential factor into a delta constraint,
\begin{align}
   \mathcal{I}^{(\alpha)}[\mathbf{y}]&= \int dt e^{it}  \tilde{\mathcal{I}}^{(\alpha)}[t,\mathbf{y}] \,, \label{eq:TIGF_Fourier}
\end{align}
where $\tilde{\mathcal{I}}^{(\alpha)}[t,\mathbf{y}]$ is defined as 
\begin{align}
	\int \prod_{j=1}^L d^D K_j&\frac{\delta( \sum_{j=1}^L\alpha_j \cdot K_j-t)\Big(\prod\limits_{k=r+1}^{s}\delta^{\lambda_{k}{-}1}(\mathcal{D}_{k})\Big)}{ \mathcal{D}_1^{\lambda_1} \cdots \mathcal{D}_r^{\lambda_r}} \, .
\end{align}
$\tilde{\mathcal{I}}^{(\alpha)}[t,\mathbf{y}]$ is a typical Feynman integrand, and conventional IBP reduction can be applied to reduce it to a set of master integrals 
\begin{align}
	\tilde{\mathcal{I}}_{1}^{(\alpha)}[t,\mathbf{y}]\,, \, \cdots \,, \, \tilde{\mathcal{I}}_{n}^{(\alpha)}[t,\mathbf{y}] \,.
\end{align}
The conventional method of evaluating master integrals using differential equations applies
\begin{align}
\begin{aligned}
\label{eq:diffeqnt}
	\partial_t \tilde{\mathcal{I}}_{i}^{(\alpha)}[t,\mathbf{y}] &= \sum_{j=1}^n A_{ij}(t,\mathbf{y}) \tilde{\mathcal{I}}_{j}^{(\alpha)}[t,\mathbf{y}] \,, \\
 \partial_{y} \tilde{\mathcal{I}}_{i}^{(\alpha)}[t,\mathbf{y}]&= \sum_{j=1}^nB_{ij}( t,\mathbf{y})\tilde{\mathcal{I}}_{j}^{(\alpha)}[t,\mathbf{y}] \,.
\end{aligned}
\end{align}

An important class of this problem is when the $t$-dependence can be solved separately from the $\mathbf{y}$-dependence, e.g. when 
\begin{align}\label{eq:factorCondition}
	A_{ij} = A_{ij}(t) \,,
\end{align} 
and spin-resummed PM dynamics falls into this class. In such a case, the $t$-dependence can be factored from the $\mathbf{y}$-dependence,
\begin{align}\label{eq:fij}
    \tilde{\mathcal{I}}_{i}^{(\alpha)}[t,\mathbf{y}]= \sum_{j=1}^nf_{ij}(t) \, \widehat{\mathcal{I}}_{j}^{(\alpha)}[\mathbf{y}] \,,
\end{align}
and the $t$ Fourier integral of \eqref{eq:TIGF_Fourier} factors from the loop integrals. The conventional differential equation approach can be applied to the second set of equations in \eqref{eq:diffeqnt} for evaluation of the effective master integrals $\widehat{\mathcal{I}}_{i}^{(\alpha)}[\mathbf{y}]$, which satisfy a modified version of the second set of equations in \eqref{eq:diffeqnt} with $B_{ij} (t, \mathbf{y})$ substituted by
\begin{align}
    \widehat{B}_{ij} (\mathbf{y}) = \sum_{i',j'=1}^{n}[f(t)^{-1}]_{ii'} B_{i'j'} (t, \mathbf{y}) f_{j'j}(t) \,,
\end{align}
where $[f(t)^{-1}]$ is the inverse matrix of $f(t)$. An observable is then given as a linear combination of the effective master integrals.

In the rest of the letter, we apply the aforementioned techniques to the evaluation of spin-resummed one-loop eikonal phase as a concrete phenomenologically relevant example.

\section{Spin-resummed 2PM eikonal} \label{sec:masterInt}
The classical eikonal can be understood as the generator of scattering observables~\cite{Gonzo:2024zxo,Kim:2024grz}, which is computed as the impact-parameter-space Fourier transform of the classical $2 \to 2$ one-loop amplitude \cite{Brandhuber:2021eyq}. The one-loop HEFT amplitude with classical spin~\cite{Chen:2024mmm} is our starting point
\begin{align}\label{eq:oneloopint}
	\begin{tikzpicture}[baseline={([yshift=-0.8ex]current bounding box.center)}]\tikzstyle{every node}=[font=\small]	
\begin{feynman}
    	 \vertex (a) {\(v_1\)};
    	 \vertex [right=1.2cm of a] (f2) [HV]{H};
    	 \vertex [right=1.cm of f2] (c){};
    	 \vertex [above=1.6cm of a](ac){$v_2$};
    	 \vertex [right=0.8cm of ac] (ad) [dot]{};
    	 \vertex [right=0.7cm of ad] (f2c) [dot]{};
    	  \vertex [above=1.6cm of c](cc){};
    	  \vertex [above=0.8cm of a] (cutL);
    	  \vertex [right=2.0cm of cutL] (cutR);
    	  \vertex [right=0.4cm of ad] (att);
    	  \vertex [above=0.3cm of att] (cut20){$ $};
    	  \vertex [below=0.3cm of att] (cut21);
    	  \diagram* {
(a) -- [fermion,thick] (f2)-- [fermion,thick] (c),
    	  (f2)--[photon,ultra thick,momentum=\(\ell_1\)](ad), (f2)-- [photon,ultra thick,momentum'=\(\ell_2\)] (f2c),(ac) -- [fermion,thick] (ad)-- [fermion,thick] (f2c)-- [fermion,thick] (cc), (cutL)--[dashed, red,thick] (cutR), (cut20)--[ red,thick] (cut21)
    	  };
    \end{feynman}  
    \end{tikzpicture}
    \begin{split}
    &M_{\rm HEFT}(q,v_1, v_2,a_1,a_2)={-1\over 2!}\\
    & \int {d^D\ell_1\over (2\pi)^{D-1} } \Big({\delta(2m_2\ell_1\mdot v_2)\over \ell_1^2\ell_2^2 } M_{a_2}(\ell_1,v_2)\\
    & \phantom{as} \times  M_{a_2}(\ell_2,v_2) \, M_{a_1}(\ell_1,\ell_2,v_2) \Big) \,,
    \end{split}
\end{align}
where $\ell_2=q-\ell_1$. $M_{a_2}$ and $M_{a_1}$ are the three-point and four-point HEFT amplitudes provided in the {\it Supplemental material}.

The key ingredient is the HEFT Compton amplitude~\cite{Bjerrum-Bohr:2023jau,Bjerrum-Bohr:2023iey}, which\textemdash contrary to the minimal coupling Compton amplitude~\cite{Arkani-Hamed:2017jhn}\textemdash is free of unphysical singularities and contains contact terms to all orders in spin, both of which contribute to the eikonal phase. However, the amplitude has spurious singularities entering through the entire functions
\begin{align}\label{eq:masterG}
G_1(x_{1}) G_1(x_{2})&= {\sinh(x_{1})\over x_{1}} {\sinh(x_{2})\over x_{2}} \,, \\
	G_2(x_{1}, x_{2})&= {1\over x_{2}}\Big({\sinh(x_{12})\over x_{12}}-\cosh(x_{2}) {\sinh(x_1)\over x_1}\Big) \,,\nn
\end{align}
where $x_{12} = x_1 + x_2$ and $x_i=a_1\mdot \ell_i$. 
\begin{figure}
    \centering
    \includegraphics[width=0.33\linewidth]{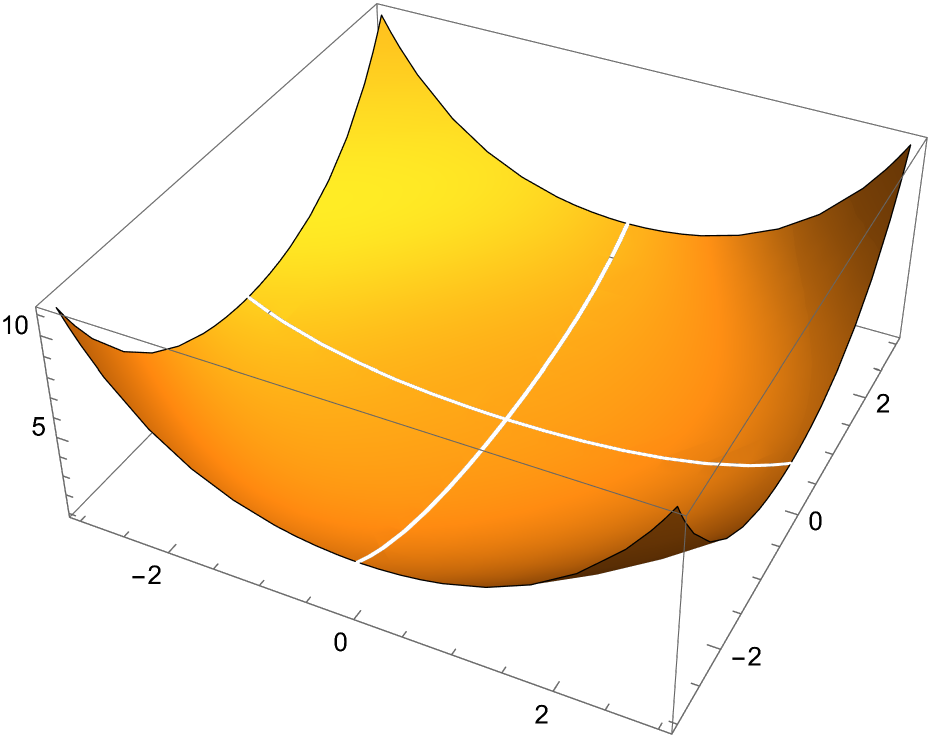}~~~~~~~~~~~~
    \includegraphics[width=0.33\linewidth]{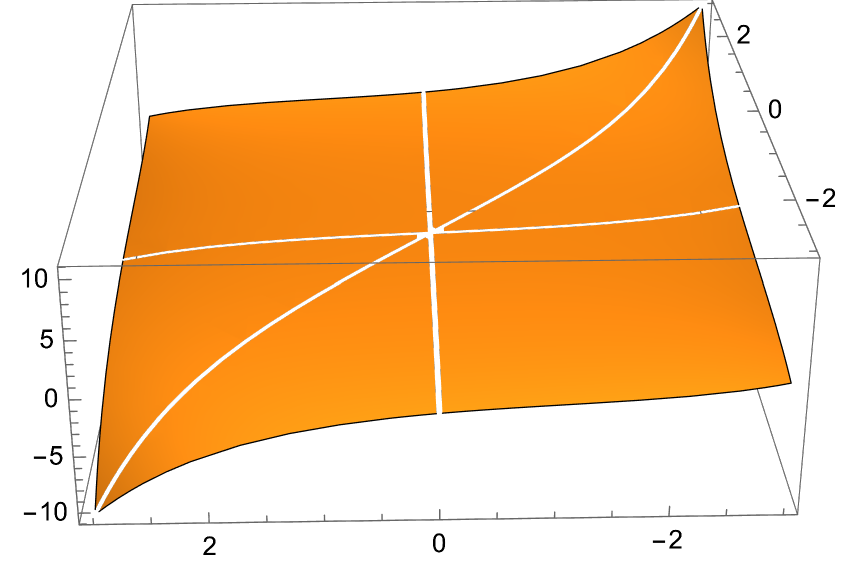}
    \caption{$G_1(x_1)G_1(x_2)$ (left) and $G_2(x_1,x_2)$ (right)}
    \label{fig:G2fun}
\end{figure}
As can be seen from the graph of the two functions in fig.~\ref{fig:G2fun}, the apparent singularities of \eqref{eq:masterG} do not correspond to actual singularities. Nevertheless, the spurious singularities can be problematic in the evaluation of the integrals, especially when the integrations are performed numerically as the singularities spoil numerical stability. We therefore introduce their integral representations
\begin{align} \label{eq:G_int_rep}
G_1(x_1)G_1(x_2)&=\int_0^1 d\sigma_1 d\sigma_2 \cosh(\sigma_1x_1) \cosh(\sigma_2x_2) \,,\nn\\
G_2(x_1,x_2)&=\int_{0}^1d\sigma_1d\sigma_2\Big( \sigma_1 \sinh \left(\sigma_1 x_1+\sigma_1 \sigma_2 x_2\right)\nn\\
&~~~~~~~~-\sinh \left(\sigma_2 x_2\right) \cosh \left(\sigma_1 x_1\right)\Big) \,,
\end{align}
which removes spurious denominators at the price of auxiliary integrations over $\sigma_i$. Analogous integral representations are used for $G_1$ functions appearing in the three-point amplitudes when constructing the integrand \eqref{eq:oneloopint}.

The eikonal $\chi$ is obtained by transforming the HEFT amplitude to impact parameter space 
\begin{align}
\begin{aligned}
    \chi &= \int {d^Dq\over (2\pi)^{D-2} } {e^{iq\mdot b} \delta(q\mdot v_1)\delta(q\mdot v_2)\over 4 m_1 m_2}
    \\ &\phantom{=asdfasdf} \times M_{\rm HEFT}(q, v_1, v_2,a_1, a_2) \,.
\end{aligned}
\end{align}
We focus on a slight generalisation of the aligned spin configuration of a binary Kerr system; $a^\mu \equiv a_1^\mu =\xi a_2^\mu$ and $v_1 \mdot a_2 = v_2 \mdot a_1=0$ with $a \mdot b \neq 0$. 

Exchanging the order of integration to pull out the $\sigma_i$ integrals, the eikonal can be schematically written as
\begin{align}\label{eq:HEFTph_sch}
   & \chi= \int_{0}^1\prod_{j=1}^4d\sigma_j \sum_{\alpha} \mathcal{I}^{(\alpha)} [\mathbf{y}] \,, 
\end{align}
where
\begin{align} \label{eq:y_def}
    y_1&=v_1\mdot v_2\,, & y_2&={ a\mdot  a \over -  b\mdot b}\,, &y_3&= b\mdot b\,, &y_4&={ b\mdot a \over - b\mdot b} \,
\end{align}
are the Lorentz scalars the eikonal depends on. The integral representation \eqref{eq:G_int_rep} reorganises the original integrand as a sum over integrands of the form \eqref{eq:TIGF_def} with additional numerator factors, which we classify into distinct \emph{sectors} $(\alpha)$ defined by the exponential factor $\exp (i q \mdot \hat b+i \ell_1 \mdot \hat a )$.
\begin{align}
	(\alpha) &\equiv(\hat b \,,\, \hat a) \\
 &\equiv(b+c'_1(\sigma) \tilde a_1+c'_2(\sigma) \tilde a_2 \,,\, c'_3(\sigma) \tilde a_1+c'_4(\sigma) \tilde a_2) \,, \nn \\
(\mathbf{y}) &=(y_1 \,,\, \hat y_2 \,,\, \hat y_3 \,,\, \hat y_4 \,,\, \sigma_1 \,,\, \sigma_2 \,,\, \sigma_3 \,,\, \sigma_4) \,, \nn \\
(K_1, K_2) &= (q \,,\, \ell_1) \,. \nn
\end{align}
The Lorentz scalars $\hat{y}_i$ are defined similar to \eqref{eq:y_def} with hatted variables $\hat b, \hat a$. The tilded spin vectors $\tilde a_j^\mu = i a_j^\mu $ are analytic continuation of the real spin(-length) vectors of the black holes $a_j^\mu$. After evaluation, the master integrals can be analytically continued back to real spin.

\begin{figure}
    \begin{tikzpicture}[baseline={([yshift=-0.8ex]current bounding box.center)}]\tikzstyle{every node}=[font=\small]	
\begin{feynman}
    	 \vertex (a)[dot] {};
        \vertex [right =1.5cm of a] (b) []{};
        \vertex [left =0.1cm of b] (bp) []{};
        \vertex [right =1.5cm of b] (c) [dot]{};
        \vertex [above =0.7cm of b] (bu) [dot]{};
         \vertex [below =0.7cm of b] (bd) [dot]{};
         \vertex [above =0.5cm of bp] (bpu) []{};
         \vertex [below =0.5cm of bp] (bpd) []{};
    	  \diagram* {
(a) -- [edge label=$\lambda_3$, ultra thick] (bu)-- [edge label=$\lambda_5$, ultra thick] (c);(a) -- [edge label'=$\lambda_1$,thick] (bd)-- [edge label'=$\lambda_4$, ultra thick] (c); (bpu) -- [edge label'=$\lambda_6$, scalar,thick] (bpd); (bu) -- [edge label=$\lambda_2$,thick] (bd);
    	  };
    \end{feynman}  
    \end{tikzpicture} ~~~~~~
    \begin{tikzpicture}[baseline={([yshift=-0.8ex]current bounding box.center)}]\tikzstyle{every node}=[font=\small]	
\begin{feynman}
    	 \vertex (a)[dot] {};
        \vertex [right =1.5cm of a] (b) []{};
        \vertex [left =0.1cm of b] (bp) []{};
        \vertex [right =1.5cm of b] (c) []{};
        \vertex [above =0.7cm of b] (bu) [dot]{};
         \vertex [below =0.7cm of b] (bd) [dot]{};
          \vertex [above =0.7cm of c] (cu) [dot]{};
         \vertex [below =0.7cm of c] (cd) [dot]{};
          \vertex [above =0.5cm of bp] (bpu) []{};
         \vertex [below =0.5cm of bp] (bpd) []{};
    	  \diagram* {
(a) -- [edge label=$\lambda_3$,ultra thick] (bu)-- [edge label=$\lambda_5$, ultra thick] (cu);(a) -- [edge label'=$\lambda_1$,thick] (bd)-- [edge label'=$\lambda_4$, ultra thick] (cd); (bpu) -- [edge label'=$\lambda_6$, scalar,thick] (bpd); (bu) -- [edge label=$\lambda_2$,thick] (bd); (cu) -- [edge label=$\lambda_8$,thick] (cd);
    	  };
    \end{feynman}  
    \end{tikzpicture}
     \caption{The graph topologies of master integrals, edges labelled by $\lambda_r$ defined in \eqref{eq:sector_defs}. The dashed line is the delta constraint originating from the exponential factor. The thick lines are remaining delta constraints, while the thin lines are physical propagators.}\label{fig:topos}
\end{figure}
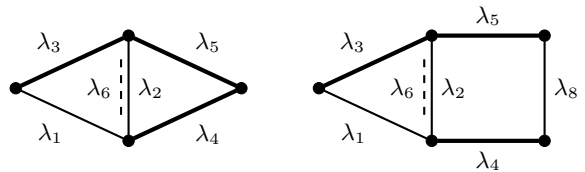

After employing the methods developed in the previous section, a typical integrand for a given sector $(\alpha)$ takes the form
\begin{align}\label{eq:IGen}
   &\mathcal{I}^{(\alpha)} [\mathbf{y}] = \int_{-\infty}^{\infty}dt e^{it}\int {d^Dq d^D\ell_1\over (2\pi)^{2D-3}} \delta(q\mdot \hat{b} + \ell_1 \mdot \hat{a} - t) \\
   &\times {\delta(q\mdot v_1)\delta(q\mdot v_2)\delta(\ell_1\mdot v_2)\over 8m_1m^2_2 \, \ell_1^2(q-\ell_1)^2}\Big(\sum_{r=0}^{2}{N^{(r)}(\mathbf{y})\over  (\ell_1\mdot v_1)^r}+{N^{(3)}(\mathbf{y})\over  q^2}\Big) \,, \nn
\end{align}
where denominators of the integrand were kept explicit. The relevant topologies are given in fig.~\ref{fig:topos}. Each sector can be reduced to master integrals by applying conventional multiloop integration techniques such as IBP reduction.

All master integrals satisfy the factorisation condition \eqref{eq:factorCondition}, where the $A_{ij} (t)$ is diagonalised. Moreover, solving the differential equations in $t$ yields power-law $t$-dependence of the master integrals. All terms resulting from reducing \eqref{eq:IGen} have $t$-dependence of the form 
\begin{align}
	{t^{j-4\epsilon}\over \epsilon} \,, && j\in [0,6] \,. 
\end{align}
In this case, the $f (t)$ matrix of \eqref{eq:fij} is diagonal and its $t$ Fourier transform can be treated as a constant factor absorbed by the effective master integrals $\widehat{\mathcal{I}}_i^{(a)} [\mathbf{y}]$. It is remarked that the divergent $\frac{1}{\epsilon}$ term in dimensional regularisation vanishes under the $t$ Fourier transform, therefore the constant factor is finite.

An exemplary term in the $N^{(3)}$ category is
\begin{align} \label{eq:I_ex}
  &\mathcal{I}_{ex}[\mathbf{y}]= \int {d^Dq d^D\ell_1\over (2\pi)^{2D-3}} e^{iq\mdot b} {\delta(q\mdot v_1)\delta(q\mdot v_2)\delta(\ell_1\mdot v_2)\over \ell_1^2(q-\ell_1)^2 q^2} \\
  & \times \frac{(16 \pi G_N)^2 m_1 m_2^2 \left(\ell _1\mdot v_1\right){}^2 \cosh \left((1-\xi)  a_1\mdot q+2 \xi  a_1\mdot \ell _1\right)}{16} \,. \nn
\end{align}
The $\cosh$ function splits the integrand into two sectors; sector $(1) = (\hat a_1 , \hat b_1) = ( 2 \tilde{a}_1 \xi \,,\, \tilde{a}_1 (1-\xi )+b)$ and sector $(2) = ( \hat a_2 , \hat b_2) = ( -2 \tilde{a}_1 \xi \,,\, \tilde{a}_1 (\xi-1 )+b )$.
Each sector reduces to three master integrals by IBP reduction,
\begin{align}
	&\mathcal{I}^{(\alpha)}_{ex}[\mathbf{y}] =-(16 \pi G_N)^2{\pi\over 2}\frac{m_1 m_2^2 (y_1^2-1) }{32 \hat y_2\hat y_3} \\
	& \phantom{as} \times \Big(2\hat y_3\left(\hat y_2+2 \hat y_4-1\right) \widehat{\mathcal{I}}_1^{(\alpha)}[\mathbf{y}]+2 \widehat{\mathcal{I}}_2^{(\alpha)}[\mathbf{y}]+\widehat{\mathcal{I}}_3^{(\alpha)}[\mathbf{y}]\Big) \,, \nn
\end{align}
where the master integrals are defined by \eqref{eq:MI_defs} and the superscript $(\alpha)$ denotes the sectors.
\eqref{eq:I_ex} becomes 
\begin{align}\label{eq:ex}
	\mathcal{I}_{ex}[\mathbf{y}]=\mathcal{I}^{(1)}_{ex}[\mathbf{y}]+\mathcal{I}^{(2)}_{ex}[\mathbf{y}] \,,
\end{align}
where each $\mathcal{I}^{(\alpha)}_{ex} [\mathbf{y}]$ is complex but the sum $\mathcal{I}_{ex} [\mathbf{y}]$ is real.

The integrals in \eqref{eq:HEFTph_sch} reduce to 320 sectors. Each sector has five master integrals. However, only three master integrals per sector appear in the final result,
 \begin{align}\label{eq:Result}
 \chi&=\int_{0}^1\prod_{j=1}^4d\sigma_j  \sum_{\alpha=1}^{320} \Big(c_{1,\alpha}(\mathbf{y}) \widehat\CI^{(\alpha)}_1[\mathbf{y}]+c_{2,\alpha}(\mathbf{y}) \widehat\CI^{(\alpha)}_2[\mathbf{y}]\nn\\
 &~~~~~~~~~~~~~~~~~~~~~~~~~~~~+c_{3,\alpha}(\mathbf{y}) \widehat\CI^{(\alpha)}_3[\mathbf{y}] \Big) \,,
 \end{align}
where the master integrals are   
 \begin{align}
\widehat\CI^{(\alpha)}_1[\mathbf{y}]&={D-4\over t^{2D-8}}J^{(\alpha)}_{1,1,1,1,1,1,0,0} \,, \nn\\
\widehat\CI^{(\alpha)}_2[\mathbf{y}]&={1\over t^{2D-10}}J^{(\alpha)}_{2,1,1,1,1,1,0,0} \,, \nn\\
\widehat\CI^{(\alpha)}_3[\mathbf{y}]&={1\over t^{2D-10}}J^{(\alpha)}_{1,1,1,1,1,1,0,1} \,, \label{eq:MI_defs}
\end{align}
defined as
\begin{align}
& J^{(\alpha)}_{\lambda_1,\lambda_2,\lambda_3,\lambda_4,\lambda_5,\lambda_6,\lambda_7,\lambda_8} \equiv \int {d^Dq d^D\ell_1\over (2\pi)^{2D-3}} \delta^{\lambda_6-1}(\hat a\mdot \ell_1+\hat b\mdot q-t)\nn\\
& \phantom{asdfasdf} \times {\delta^{\lambda_3-1}(\ell_1\mdot v_2)\delta^{\lambda_4-1}(q\mdot v_1)\delta^{\lambda_5-1}(q\mdot v_2)\over(\ell_1^2)^{\lambda_1}((q-\ell_1)^2)^{\lambda_2} (\ell_1\mdot v_1)^{\lambda_7}(q^2)^{\lambda_8}} \,. \label{eq:sector_defs}
\end{align}
The numerators $N^{(0,1,2)}$ of \eqref{eq:HEFTph_sch} reduce to $\widehat\CI^{(\alpha)}_{1,2}[\mathbf{y}]$, while the numerators $N^{(3)}$ of \eqref{eq:HEFTph_sch} reduce to $\widehat\CI^{(\alpha)}_{1,2,3}[\mathbf{y}]$. The integrals $\widehat\CI^{(\alpha)}_{1,2}[\mathbf{y}]$ correspond to the left topology of fig.~\ref{fig:topos}, while the integral $\widehat\CI^{(\alpha)}_{3}[\mathbf{y}]$ corresponds to the right topology of fig.~\ref{fig:topos}. The apparent $t$-dependence in \eqref{eq:MI_defs} makes the master integrals $\widehat\CI^{(\alpha)}_{1,2,3}[\mathbf{y}]$ independent of $t$. As mentioned previously, the other two master integrals, $J^{(\alpha)}_{1,1,1,2,1,1,0,0}$ and $J^{(\alpha)}_{1,1,1,1,1,1,1,0}$, never appear and their evaluation is unnecessary for evaluating the eikonal. It is also remarked that both $c_{j,\alpha}(\mathbf{y})$ and $\widehat\CI^{(\alpha)}_j[\mathbf{y}]$ are finite in dimensional regularisation. 

The final result for the eikonal phase is available in the public repository  {\href{https://github.com/AmplitudeGravity/KerrEikonal2pm}{\blue \texttt{KerrEikonal2pm}}}~\cite{ChenGitHub}. We emphasise that the result contains all-orders-in-spin contributions for the generalised aligned spin configuration with $a\mdot b\neq 0$.   

\section{Evaluation of the Master integrals} 

We evaluate the master integrals using the method of differential equations~\cite{gehrmann2000differential,Kotikov:1990kg,Henn:2013pwa}.  The $\hat{y}_3$-dependence of the master integrals can be determined from dimensional analysis, since the only dimensionful variable is $\hat{y}_3$. Moreover, the differential operator $\partial_{y_1}$ is diagonal in the system of master integrals, and the $y_1$-dependence of the master integrals can be solved separately.

The nontrivial systems of differential equations are given by $\hat{y}_{2,4}$. The two systems are degenerate for the master integrals $\widehat{\mathcal{I}}_{1,2}$ and they only depend on the combination $\hat{y}_2' = \hat{y}_2 + 2 \hat{y}_4$. 
The differential equations over $\hat{y}_2'$ for $\widehat{\mathcal{I}}_{1,2}$ are solved by complete elliptic integrals~\footnote{We adopt \texttt{Mathematica}'s definition for the elliptic integrals.},
\begin{align}
    \widehat{\mathcal{I}}_1[\mathbf{y}]&=\frac{C}{\sqrt{-\hat{y}_3}\sqrt{y_1^2-1}} \, \frac{K(\hat{y}_2')}{\pi }\,, \nn\\
    \widehat{\mathcal{I}}_2[\mathbf{y}]&= -   \frac{C\sqrt{-\hat{y}_3}}{\sqrt{y_1^2-1}} \, \frac{(\hat{y}_2'-1) K(\hat{y}_2') + E( \hat{y}_2')}{\pi } \,,
\end{align}
where the integration constant $C = -\frac{1}{16 \pi^2}$ is fixed from the spinless limit $\hat{y}_2' \to 0$.

The differential equations satisfied by $\widehat{\mathcal{I}}_3$ are 
\begin{align}
&\partial_{\hat y'_2}\widehat{\mathcal{I}}_3[\mathbf{y}]={-1\over 2 \left(\hat y'_2-2 \hat y_4\right) \left(\hat y'_2+\hat y^2_4 -2\hat y_4\right)}\times\nn\\
    &\Big(\hat y_4^2 \widehat{\mathcal{I}}_3[\mathbf{y}]-\frac{C}{\pi}\frac{(\sqrt{- \hat{y}_3}) }{\sqrt{y_1^2-1}} \left(\hat y'_2+2 \hat y_4 \left(\hat y_4-1\right)\right) E\left(\hat y'_2\right)\nn\\
    &+\frac{C}{\pi}\frac{(\sqrt{- \hat{y}_3}) }{\sqrt{y_1^2-1}}\left(1-\hat y_4\right) \left(\hat y'_2-2 \hat y_4\right) K\left(\hat y'_2\right)\Big)\, , \label{eq:diffI5_1} \\
& \partial_{\hat y_4}\widehat{\mathcal{I}}_3[\mathbf{y}]={1\over \left(\hat y'_2-2 \hat y_4\right) \left(\hat y'_2+\hat y^2_4 -2\hat y_4\right)} \Big(\hat y_4 \left(\hat y'_2-\hat y_4\right) \widehat{\mathcal{I}}_3[\mathbf{y}]\nn\\
&
     -C \frac{(\sqrt{- \hat{y}_3}) }{\sqrt{y_1^2-1}}\left(\hat y'_2-1\right) \left(\hat y'_2-2 \hat y_4\right) K\left(\hat y'_2\right)\nn\\
     & -C \frac{(\sqrt{- \hat{y}_3}) }{\sqrt{y_1^2-1}}\left(\hat y_4 \hat y'_2+\hat y'_2-2 \hat y_4\right) E\left(\hat y'_2\right)\Big) \, . \label{eq:diffI5_2}
\end{align}
The common denominator factor in the two equations implies that the master integral should be understood as an integral over the elliptic curve
\begin{align}\label{eq:eCurve}
   \CY^2 = (  \hat{y}'_2-2z ) (z^2 - 2z + \hat{y}'_2) \,,
\end{align}
where $z$ is the integration variable to be identified as $ - \hat{y}_4$ after integration. $\widehat{\mathcal{I}}_{3} [\mathbf{y}]$ is computed by solving \eqref{eq:diffI5_1} at $\hat{y}_4 = 0$, which yields the boundary value for the remaining integration over $\hat{y}_4$. Carrying out the $\hat{y}_4$ integration, we have
\begin{align}
  \widehat{\mathcal{I}}_3[\mathbf{y}] &=  \frac{C(\sqrt{- \hat{y}_3}) }{\sqrt{y_1^2-1}}\frac{\sqrt{\hat y'_2+\hat y_4 \left(\hat y_4-2\right)} }{\sqrt{\hat y'_2-2 \hat y_4}}\Bigg[\frac{ 2E(\hat y'_2)+\pi }{2\pi} \nn\\
  &\phantom{=a} + \int _0^{\hat y_4} dz \Big(\frac{ \left(1-\hat y'_2\right) \left(\hat y'_2-2 z\right)K\left(\hat y'_2\right)}{\pi\,  \CY \, ( z^2-2z+\hat y'_2)}
  \\ &\phantom{=asdfasdfasdf} - \frac{ \left(\hat y'_2 z-2z+\hat y'_2\right) E\left(\hat y'_2\right)}{\pi\,  \CY \, ( z^2-2z+\hat y'_2)} \Big)\Bigg] \,. \nn
\end{align}
The $z$-integral can be reduced by the IBP relations following from the exact differentials
\begin{align}
    d \left( \frac{\hat{y}'_2-2 z}{\CY} \right) \,,\, d \left( \frac{(\hat{y}'_2 - 2 z )^2}{\CY} \right) \,.
\end{align}
Eliminating integrals with undesired denominators, we arrive at
\begin{align}\label{eq:I5}
	\widehat{\mathcal{I}}_3[\mathbf{y}] &= \, \frac{C\sqrt{- \hat{y}_3} }{\sqrt{y_1^2-1}} \Bigg[\frac{\sqrt{\hat{y}'_2 + \hat{y}_4^2 - 2 \hat{y}_4} }{\pi  \sqrt{\hat{y}'_2 - 2 \hat{y}_4}} \Big({\pi-2K(\hat{y}'_2)\over 2}\nn\\
	& - \left(E(\hat{y}'_2)-K(\hat{y}'_2) \right) \int _0^{\hat{y}_4} \frac{dz}{\CY} - K(\hat{y}'_2) \int _0^{\hat{y}_4} \frac{zdz}{\CY} \Big) \nn\\
	&  +  \, \frac{E(\hat{y}'_2)+\left(1-\hat{y}_4 \right) K(\hat{y}'_2)}{\pi } \Bigg]\,,
\end{align}
where the integrals can be converted to incomplete elliptic integrals, which can be evaluated to arbitrary numerical precision. The integration constants were fixed by requiring regularity of the spin expansion around $\hat{a}^\mu = 0$. Since $\hat{a}^\mu \to 0$ is a singular limit of the underlying curve \eqref{eq:eCurve}, we expand \eqref{eq:I5} around $\hat{y}_4$ first and then expand in $\hat{y}_2$ to obtain the series expansion of $\widehat{\mathcal{I}}_3$. The master integrals $\widehat{\mathcal{I}}_{1,2,3}$ were checked against brute-force calculations based on the formulae in appendix B of ref.~\cite{Kim:2024grz}, and found to agree up to $\mathcal{O}(\hat{a}^{20})$.

\section{Results}
The eikonal phase \eqref{eq:Result} is exact to all orders in spin, and can be expanded in spin before the $\sigma_j$-integration for a check against perturbative-in-spin calculations. 
The spin-expanded $\sigma_j$ integrand turns out to be polynomial in $\sigma_j$, and the auxiliary $\sigma_j$ integrals can be evaluated exactly. We provide files for evaluating the spin-expanded $\sigma_j$ integrand in the public repository {\href{https://github.com/AmplitudeGravity/KerrEikonal2pm}{\blue \texttt{KerrEikonal2pm}}}~\cite{ChenGitHub}. The spin-expanded result up to $\mathcal{O}(a^8)$ order is included in the repository and have been checked to be consistent with the results of ref.~\cite{Chen:2024mmm}.

\begin{figure}
    \centering
    \begin{minipage}[c]{0.2\textwidth}
    	\includegraphics[width=1\linewidth]{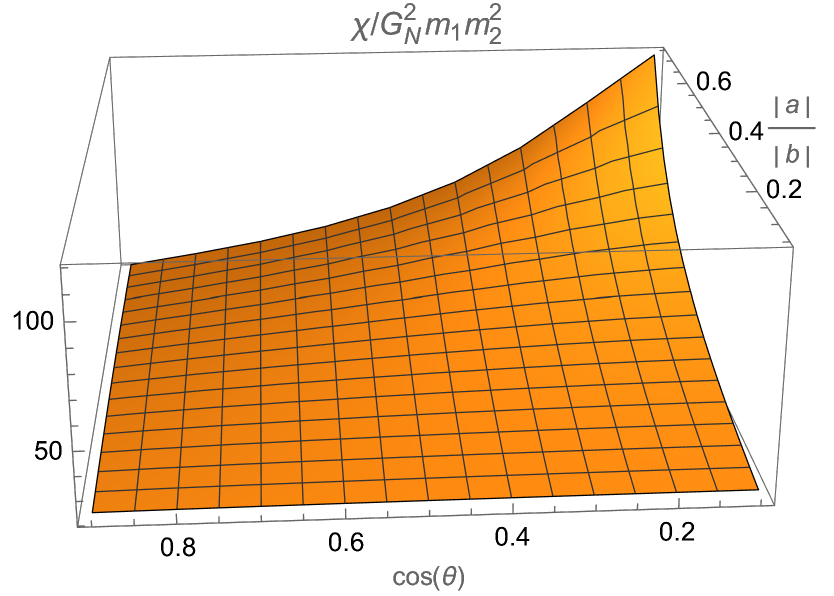}
    	\caption{The eikonal \eqref{eq:Result} 
    as a function of ${|a|\over |b|}$ and $\cos(\theta) = \frac{-a\cdot b}{|a||b|}$ with numerical errors $\sim 10^{-15}$.}
     \label{fig:heftphase}
    \end{minipage}
    \hfill
\begin{minipage}[t]{0.25\textwidth}
    \begin{tikzpicture}
	\begin{axis}[width=1\textwidth,
			title={},
			xmin=1.0, xmax=1.2,
			ymin=4.8, ymax=10,
			axis lines=left,
			xtick={1.0,1.05,1.10,1.15,1.2},
			ytick={5,6,7,8,9},
			compat=newest,
			xlabel=$|b|/|a|$, 
			ylabel= $\log(\frac{\chi}{G_N^2 m_1 m_2^2})$, ylabel style={rotate=0,at={(-0.07,0.7)},anchor=south},
			legend pos=north east,
			ymajorgrids=true,
			grid style=dashed,
            every axis plot/.append style={thick},
		]		
		\addplot[
			color=blue,
			mark=star,
		]
		coordinates {(1.01,9.4973813516)(1.02,8.474859211)(1.03,7.8853231666)(1.04,7.471039967)(1.05,7.156824413)(1.06,6.901244374)(1.07,6.6870761397)(1.08,6.503174376)(1.09,6.344109763)(1.1,6.209525958)(1.11,6.080087102)(1.12,5.964705154)(1.13,5.861490680)(1.14,5.764372280)(1.15,5.678363889)(1.16,5.595842996)(1.17,5.519000526)(1.18,5.449393789)(1.19,5.382457651)(1.2,5.319755193)};
		\addplot[
			color=orange,
			mark=none,
		]
		coordinates {(1.01,9.52411)(1.02,8.47695)(1.03,7.86134)(1.04,7.42245)(1.05,7.0804)(1.06,6.79962)(1.07,6.56113)(1.08,6.3536)(1.09,6.16973)(1.1,6.00453)(1.11,5.85444)(1.12,5.71683)(1.13,5.58971)(1.14,5.47152)(1.15,5.36104)(1.16,5.25727)(1.17,5.15941)(1.18,5.06677)(1.19,4.97881)(1.2,4.89503)};
		   \legend{{\rm eikonal},{\rm reference}}
	\end{axis}
\end{tikzpicture}
\caption{Aligned-spin $\chi$ near the singular point $|b| = |a|$. The curve $\frac{\beta}{(|b|^2 - |a|^2)^{3/2}}$ intersects at $|b| = 1.02 \, |a|$.}
\label{fig:scaling}
\end{minipage}
\end{figure} 
The merit of the eikonal phase \eqref{eq:Result} is that we can study its full spin dependence quantitatively, since the $\sigma_j$ integrations can be numerically evaluated with high precision. As an example of numerically studying the eikonal \eqref{eq:Result}, we consider the scattering of a test Kerr black hole on a Schwarzschild background, corresponding to the substitution
 \begin{align} \label{eq:eik_plot_cond}
    \chi_{a_2=0} = \chi|_{\xi\rightarrow 0} \,,
\end{align}
which is presented in fig.~\ref{fig:heftphase}. The parameters $y_1={12\over 10}$ and $y_3=-1$ are chosen for the ranges $|\sqrt{-y_2}| < 0.75$ and $\cos(\theta) \in [0,0.9]$, where $\theta$ denotes the angle between $a^\mu$ and $b^\mu$ such that $\sqrt{-y_2}\cos(\theta)=y_4$. 

The numerical calculations can be tested against analytic predictions. For example, the 2PM aligned-spin ($\cos(\theta)=0$) eikonal is expected to have the singularity $\propto (|b|^2 - |a|^2)^{-3/2}$ based on analytic studies~\cite{Damgaard:2022jem,Aoude:2022thd,Kim:2024grz}. Fig.~\ref{fig:scaling} shows numerical evaluation of \eqref{eq:Result} in the aligned-spin configuration plotted against the reference curve $\frac{\beta}{(|b|^2 - |a|^2)^{3/2}}$, where the constant $\beta$ was fit to \eqref{eq:Result} at $|b| = 1.02 \, |a|$ and other unspecified parameters are the same as in fig.~\ref{fig:heftphase}. The two lines on fig.~\ref{fig:scaling} show a good agreement near the singular point $|b| = |a|$, having $\lesssim 5 \, \%$ relative  difference for $|b| \le 1.04 \, |a|$. This singularity can be interpreted as the dynamics detecting the ring singularity from spin resummation, since the eikonal has a singularity only at $|b| = 0$ when spin is treated perturbatively.

\section{Conclusions and outlook}
In this letter, a systematic method to reduce and evaluate TIGFs was introduced and applied to the scattering dynamics of a binary Kerr system in the generalised aligned-spin configuration. The corresponding 2PM eikonal phase \eqref{eq:Result} was presented in a closed form, which can be studied analytically by expanding to arbitrary orders in spin, or can be studied numerically through numerical integrations for exact spin dependence.

An important application of TIGFs is generation of tensor integrals, directly providing an alternative method of performing tensor reduction in Feynman integrals~\cite{Feng:2022hyg}. The approach may prove useful for reduction of irreducible numerators, which in many cases becomes a bottleneck when evaluation of multiloop Feynman integrals are involved. In this regard, understanding the criteria for the factorisation of the $t$-dependence (e.g. from intersection theory \cite{cho1995intersection,Mastrolia:2018uzb,Frellesvig:2019kgj,Brunello:2024ibk}) will be important, as the factorisation played a crucial role in obtaining closed-form expressions for the TIGFs considered. 

Another future direction to explore would be the study of function space complexity for TIGFs. We have seen that elliptic integrals appear in effective two-loop (2PM) topologies for TIGFs (see also ref.~\cite{Kim:2024grz} for electromagnetism). It is reasonable to expect that more complex functions (e.g. elliptic multi-polylogarithms \cite{Broedel:2017kkb}, integrals over Calabi-Yau manifolds \cite{Bourjaily:2022bwx}) will appear in effective three-loop (3PM) graph topologies when integrals are deformed into TIGFs. Moreover, we can attempt to quantify the increase in transcendentality of the function space when typical Feynman integrals are deformed into TIGFs. This may even lead to an exploration of new geometries that have not yet been associated with Feynman integrals.

Special cases of TIGFs appear ubiquitously in quantum-field-theory-inspired approaches to classical gravitating systems, especially in the calculation of scattering waveforms~\cite{Brandhuber:2023hhy,Herderschee:2023fxh,Elkhidir:2023dco,Georgoudis:2023lgf,Caron-Huot:2023vxl,Bohnenblust:2023qmy,Bini:2023fiz,Georgoudis:2023eke,Adamo:2024oxy,Bini:2024rsy,Alessio:2024wmz,Georgoudis:2024pdz,Brunello:2024ibk}. It would also be interesting to apply the developed methods to these problems.

Coming back to the initial motivation for the study of TIGFs, it would be interesting to apply the developed methods to 3PM all-orders-in-spin dynamics and study spin resummation. Whether we have the exact three-graviton-Kerr five-point amplitude\textemdash which is necessary for constructing the 3PM integrand\textemdash is less relevant, as long as the conjectured five-point amplitude correctly captures features of the dynamical Newman-Janis shift; we expect the singularity structures of the binary Kerr dynamics to be governed by the dynamical Newman-Janis shift, and correct singularity structures are the most important when we attempt to resum the perturbative two-body dynamics~\cite{Kim:2024grz}. Compared to the 2PM dynamics studied in this letter, the spin-resummed 3PM dynamics is qualitatively different in that it is the first order where next-to-leading-order effects in the mass-ratio expansion enters into the dynamics~\cite{Vines:2018gqi}, thereby including the first beyond background-probe limit effects. The insights gained from studying singularity structures of spin-resummed binary Kerr dynamics may motivate new resummation schemes for spinning binary dynamics, providing a more accurate incorporation of spin effects in waveform models used by gravitational wave observatories, potentially having far-reaching consequences for astrophysics and multi-messenger astronomy.

\section*{Acknowledgements}
It is a pleasure to thank Emil Bjerrum-Bohr, Andreas Brandhuber, Graham Brown, Marcos Skowronek, Zhengwen Liu, Roger Morales, Gabriele Travaglini for interesting conversations.  JWK would like to thank Stefano De Angelis and Fei Teng for stimulating discussions. The authors would like to thank Bo Feng, Andres Luna, and Roger Morales for comments on the draft. GC has received funding from the European Union's Horizon 2020 research and innovation program under the Marie Sk\l{}odowska-Curie grant agreement No.~847523 ``INTERACTIONS''. TW is supported by the NRF grant 2021R1A2C2012350.

\bibliographystyle{apsrev4-1}
\bibliography{KinematicAlgebra6}

\appendix
\onecolumngrid
\begin{center}
    \textbf{\large Supplemental material for  \\``Systematic integral evaluation for spin-resummed binary dynamics''}
\end{center}
In this Supplemental Material we present the tree amplitude used in the main text. \\ ~ \\
\twocolumngrid
The three point amplitude with one graviton momentum $p_1^\mu$ and classical black hole spin $a^\mu$~\cite{Guevara:2018wpp,Chung:2018kqs,Arkani-Hamed:2019ymq,Aoude:2020onz,Johansson:2019dnu,Bjerrum-Bohr:2023jau} is    
\begin{align} \label{threepointgrav}
    M_3(p_1\mdot a)=  \sqrt{32 \pi G_N} \,(\pb\Cdot\varepsilon_1)(w_1\mdot\varepsilon_1)\,,
\end{align}
where $\bar p^\mu=m v^{\mu}$ denotes massive particle's momentum,
\begin{align}
\begin{split}
    w_1^\mu&\coloneqq\cosh(p_1\Cdot a) \, \pb^\mu-i\frac{\sinh(p_1\Cdot a)}{p_1\Cdot a}(p_1\Cdot S)^\mu
    %
    \, , 
    \end{split}
\end{align}
and $(p_1 \mdot S)^\mu = p_{1\nu}S^{\nu\mu}$. The $S^{\mu\nu}$ denotes the spin tensor $S^{\mu\nu}=-\epsilon^{\mu\nu\rho\sigma}\bar p_\rho a_\sigma$.

The Compton amplitude with graviton momenta $p_{1,2}$ is constructed from the double copy \cite{Bern:2008qj,Bern:2010ue}, kinematic Hopf algebra \cite{Brandhuber:2021kpo,Brandhuber:2021bsf,Brandhuber:2022enp}, and classical spin bootstrap~\cite{Bjerrum-Bohr:2023iey,Bjerrum-Bohr:2023jau},
\begin{widetext}
\begin{align}
\label{fullCompton}
    &{M_4\over 32 \pi G_N}=-\Big(\!\frac{\bar p\mdot F_1\mdot F_2\mdot \bar p}{p_2\mdot \bar p}\Big)\Bigg(\frac{w_1\mdot F_1\mdot F_2\mdot w_2}{2(p_1\mdot p_2) (p_1\mdot \bar p)}-\frac{p_1\mdot \bar p-p_2\mdot \bar p}{4(p_1\mdot p_2)(p_1\mdot \bar p)}
     \Big(i G_2\left(x_1,x_2\right) (a\mdot F_1\mdot F_2\mdot S\mdot p_2)+i G_2(x_1,x_2) (a\mdot F_2\mdot F_1 \mdot S\mdot p_1) \nn\\
     &+i G_1\left(x_{12}\right) \text{tr}\left(F_1\mdot S\mdot F_2\right)+ G_1(x_1) G_1(x_2) \big( (a\mdot F_1\mdot \bar p) (a\mdot F_2\mdot p_1)\!-\!(a\mdot F_1\mdot p_2) (a\mdot F_2\mdot \bar p) -\!\frac{p_2\mdot \bar p\!-\!p_1\mdot \bar p}{2}  (a\mdot F_1\mdot F_2\mdot a)\big)\Big)\Bigg)\, \nn\\
      &+\!{\Big((\partial_{x_1}\!-\!\partial_{x_2})G_1(x_1) G_1(x_2)\Big)\over 4(\bar p\mdot p_1) (\bar p\mdot p_2)} \Big(\bar p\mdot p_2(\bar p^2 (a\mdot F_1\mdot F_2\mdot a) (a\mdot F_2\mdot F_1\mdot \bar p) +a^2 (\bar p_4\mdot F_1\mdot F_2\mdot \bar p) (a\mdot F_1\mdot F_2\mdot \bar p))\!-\! (1\leftrightarrow 2)\!\Big)\nn\\
     &+\Big({i(\partial_{x_1}\!-\!\partial_{x_2})G_2(x_1,x_2)\over 4(\bar p\mdot p_1) (\bar p\mdot p_2)}\Big) \Big((\bar p\mdot p_2) (a\mdot F_2\mdot F_1\mdot \bar p) ((a\mdot F_2\mdot \bar p) (a\mdot \tF_1\mdot \bar p){-}(a\mdot F_1\mdot \bar p) (a\mdot \tF_2\mdot \bar p))+(1\leftrightarrow 2)\Big)\nn\\
      &+\!\Big(\!{(\partial_{x_1}\!-\!\partial_{x_2})^2\over 2!}{G_{1}(x_1)}{G_{1}(x_2)}\Big)\Big((a\mdot F_1\mdot \bar p) (a\mdot F_2\mdot \bar p) (a\mdot  F_1\mdot F_2\mdot a)\!-\!{a^2\over 2} ((a\mdot F_1 \mdot F_2\mdot p) (a\mdot F_2 \mdot F_1\mdot \bar p) - (a\mdot F_1 \mdot F_2\mdot a) (\bar p\mdot F_1 \mdot F_2\mdot \bar p))\!\Big)\nn\\
     &\!\!\!+\Big({i(\partial_{x_1}-\partial_{x_2})^2\over 2!}{G_{2}(x_1,x_2)}\Big)\Big(\! -{1\over 2} \big((a\mdot F_1\mdot F_2\mdot a) (a\mdot F_2\mdot \bar p) (a\mdot \tF_1\mdot \bar p)\!-\!(1\leftrightarrow 2)\big)\Big)\, ,
\end{align}
\end{widetext}
where $x_i=a\mdot p_i$, $F_i^{\mu\nu}= p^\mu_i\vareps^\nu_i-\vareps^\mu_ip^\nu_i$, and $\widetilde F^{\mu\nu}= \frac{1}{2} \epsilon^{\mu\nu\rho\sigma} F_{\rho\sigma}$.
The differences between this Compton amplitude and other proposals~\cite{Cangemi:2023bpe,Bautista:2023sdf} 
stem from the differences in their respective formalisms and assumptions~\cite{Bjerrum-Bohr:2023jau,Bjerrum-Bohr:2023iey}, which are inherited by the one-loop amplitude~\cite{Chen:2024mmm}. 
The recursion relation for the integral representation of the $G$ functions is
\begin{align}
G_1(x_1)&{=}\int_0^1 d\sigma_1 \cosh(\sigma_1x_1) \,, \\
G_2(x_1,x_2)&{=}\int_{0}^1d\sigma_2\Big[\partial_{x_1}G_{1}(x_1+\sigma_2 x_2) \nn \\
&~~~~~~~~~-\sinh(\sigma_2 x_2) \cosh(\sigma_1 x_1)\Big] \,, \\
G_r(x_1, ..., x_r)&{=}{\int_0^1} d\sigma_{r}\Big[\partial_{x_1}G_{r-1}(x_1{+}\sigma_r x_r, x_2,..., x_{r-1})\nn\\
& -G_{r-1}(x_1, ... , x_{r-1})\sinh(\sigma_rx_r)\Big] \, . 
\end{align}
\end{document}